\documentclass[aps,pra,reprint,nofootinbib,twocolumn,superscriptaddress,showpacs,showkeys,longbibliography,floatfix,bibnotes]{revtex4-1}
\usepackage{eurosym}
\usepackage{amsmath,amssymb,amstext}
\usepackage{ulem}
\usepackage[usenames,dvipsnames]{color}
\usepackage{bm}
\usepackage{graphicx}
\usepackage{braket}
\usepackage{natbib}
\usepackage{comment}
\usepackage{dcolumn}
\usepackage[english]{babel}
\usepackage{wasysym}
\usepackage{subfigure}
\usepackage[colorlinks,bookmarks=false,citecolor=blue,linkcolor=red,urlcolor=blue]{hyperref}

\usepackage{graphicx}  
\usepackage{float}  
\usepackage{subfigure}  
\usepackage{overpic}
\usepackage{hyperref}
\hypersetup{hypertex=true,
            colorlinks=true,
            linkcolor=blue,
            anchorcolor=blue,
            citecolor=blue}

\begin{document}

	\title{Exceptional nexus in Bose-Einstein condensates with collective dissipation}
	
	\author{Chenhao Wang}
	\thanks{These two authors contributed equally.}
	\affiliation{State Key Laboratory of Quantum Optics and Quantum Optics
		Devices, Institute of Laser Spectroscopy, Shanxi University, Taiyuan 030006,
		China} 
	\affiliation{Collaborative Innovation Center of Extreme Optics,
		Shanxi University, Taiyuan 030006, China}
	
	\author{Nan Li}
	\thanks{These two authors contributed equally.}
	\affiliation{State Key Laboratory of Quantum Optics and Quantum Optics
		Devices, Institute of Laser Spectroscopy, Shanxi University, Taiyuan 030006,
		China} 
	\affiliation{Collaborative Innovation Center of Extreme Optics,
		Shanxi University, Taiyuan 030006, China}

     \author{Jin Xie}
	\affiliation{State Key Laboratory of Quantum Optics and Quantum Optics
		Devices, Institute of Laser Spectroscopy, Shanxi University, Taiyuan 030006,
		China} 
	\affiliation{Collaborative Innovation Center of Extreme Optics,
		Shanxi University, Taiyuan 030006, China}
		
		\author{Cong Ding}
	\affiliation{State Key Laboratory of Quantum Optics and Quantum Optics
		Devices, Institute of Laser Spectroscopy, Shanxi University, Taiyuan 030006,
		China} 
	\affiliation{Collaborative Innovation Center of Extreme Optics,
		Shanxi University, Taiyuan 030006, China}
		
		\author{Zhonghua Ji}
	\affiliation{State Key Laboratory of Quantum Optics and Quantum Optics
		Devices, Institute of Laser Spectroscopy, Shanxi University, Taiyuan 030006,
		China} 
	\affiliation{Collaborative Innovation Center of Extreme Optics,
		Shanxi University, Taiyuan 030006, China}

	\author{Liantuan Xiao}
	\affiliation{State Key Laboratory of Quantum Optics and Quantum Optics
		Devices, Institute of Laser Spectroscopy, Shanxi University, Taiyuan 030006,
		China} 
	\affiliation{Collaborative Innovation Center of Extreme Optics,
		Shanxi University, Taiyuan 030006, China}
	
	\author{Suotang Jia}
	\affiliation{State Key Laboratory of Quantum Optics and Quantum Optics
		Devices, Institute of Laser Spectroscopy, Shanxi University, Taiyuan 030006,
		China} 
	\affiliation{Collaborative Innovation Center of Extreme Optics,
		Shanxi University, Taiyuan 030006, China}
		
\author{Bo Yan}
	\affiliation{{Zhejiang Key Laboratory of Micro-nano Quantum Chips and Quantum Control, School of Physics, and State Key Laboratory for Extreme Photonics and Instrumentation, Zhejiang University, Hangzhou 310027, China}} 

 \author{Ying Hu}
	\thanks{huying@sxu.edu.cn}
	\affiliation{State Key Laboratory of Quantum Optics and Quantum Optics
		Devices, Institute of Laser Spectroscopy, Shanxi University, Taiyuan 030006,
		China} 
	\affiliation{Collaborative Innovation Center of Extreme Optics,
		Shanxi University, Taiyuan 030006, China}

	\author{Yanting Zhao}
	\thanks{zhaoyt@sxu.edu.cn}
	\affiliation{State Key Laboratory of Quantum Optics and Quantum Optics
		Devices, Institute of Laser Spectroscopy, Shanxi University, Taiyuan 030006,
		China} 
	\affiliation{Collaborative Innovation Center of Extreme Optics,
		Shanxi University, Taiyuan 030006, China}

\begin{abstract}
In multistate non-Hermitian systems, higher-order exceptional points and exotic phenomena with no analogues in two-level systems arise. A paradigm is the exceptional nexus (EX), a third-order EP as the cusp singularity of exceptional arcs (EAs), that has a hybrid topological nature. Using atomic Bose-Einstein condensates to implement a dissipative three-state system, we experimentally realize an EX within a two-parameter space, despite the absence of symmetry. The engineered dissipation exhibits density dependence due to the collective atomic response to resonant light. Based on extensive analysis of the system's decay dynamics, we demonstrate the formation of an EX from the coalescence of two EAs with distinct geometries. These structures arise from the different roles played by dissipation in the strong coupling limit and quantum Zeno regime. Our work paves the way for exploring higher-order exceptional physics in the many-body setting of ultracold atoms. 
\end{abstract}

\maketitle	

The exceptional point (EP), a branch point singularity in the spectrum, is at the heart of many fascinating non-Hermitian phenomena and applications with no Hermitian counterparts~\cite{Heiss2012,Ashida2020,Bergholtz2021,Ding2022}. At an EP of order $N$, $N$ eigenvalues and the corresponding eigenstates of a non-Hermitian Hamiltonian simultaneously coalesce. The simplest is the second-order EP (EP2) of a two-state non-Hermitian Hamiltonian, which has been extensively studied in experiments~\cite{Bergholtz2021,Ding2022}. Lately, observations of EP2s in quantum systems were reported, including atoms~\cite{Peng2016,Li2019,Ren2022,Liang2023}, ion traps~\cite{DingL2021}, single spins~\cite{Wu2019} and cavities~\cite{Choi2010}.

In non-Hermitian systems with more than two states, multiple EPs can form and interact, resulting in qualitatively novel phenomenology absent in two-state cases~\cite{Graefe2008,Demange2012,Ding2016,Xiao2019,Tang2020,Mandal2021,Delplace2021, Bergholtz2021,Ding2022}. Each trajectory of these EPs can trace out interesting geometries in the parameter space, yielding a kaleidoscope of arcs, rings, etc. Their interactions further lead to the coalescence of EPs and creation of higher-order EPs, which entail novel physics.  A paradigm is where the coalescence of multiple exceptional arcs (EAs) consisting of EP2s produces an exceptional nexus (EX)~\cite{Ding2016,Xiao2019,Tang2020}, which is not only a third-order EP (EP3) but also the cusp singularity of EAs. Different from an EP2, which has a half topological charge, an EX has a unique, hybrid topological nature associated with two distinct topological invariants~\cite{Xiao2019,Tang2020}. 
\begin{figure}
 
\includegraphics[width=1\columnwidth]{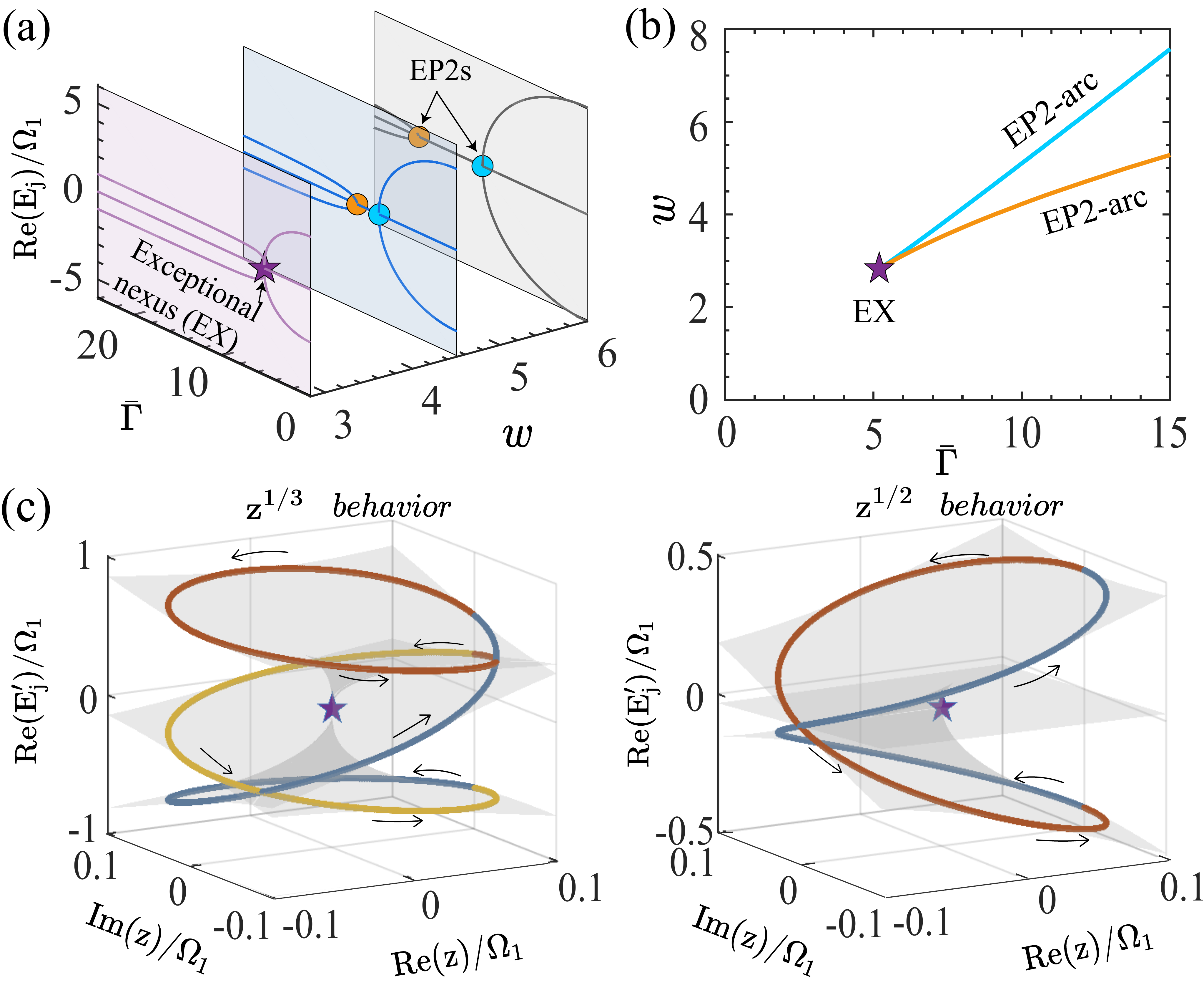} 
\vspace{-0.5cm}  
 \caption{Exceptional nexus (EX) formed in two-dimensional (2D) parameter space in the absence of symmetry. (a) Calculated eigenvalues of Hamiltonian~(\ref{H}) as a function of dissipation rate $\bar{\Gamma}$, for $w=6, 4.5, 2\sqrt{2}$ (grey, blue and purple), respectively. Only the real components are shown. (b) Two EAs with different EP geometries merge at the EX in the $(\bar{\Gamma},w)$ plane. (c) Anisotropic perturbation effects near the EX. We numerically calculate the eigenvalues $E^{\prime}_j$ ($j=1,2,3$) and eigenstates of the perturbed Hamiltonian $H'=H_\textrm{EX}+zH_1$, where $z/\Omega_1= 0.1 e^{i\theta}$. Left panel: $H_1=|3\rangle\langle 3|$. Right panel: $H_1=i|1\rangle\langle 3|-i|3\rangle\langle 1|+\frac{\sqrt{3}}{3}|2\rangle\langle 1|+\frac{\sqrt{6}}{6}|2\rangle\langle 3|$. Real parts of $E^{\prime}_j$ are shown as a function of $z$.}  \label{Fig1}
\vspace{-0.7cm}
\end{figure}

Nonetheless, experimental explorations of higher-order exceptional phenomena pose challenges due to the need to tune more parameters: An order-$N$ EP generically requires the tuning of $2(N-1)$ real parameters~\cite{Pan2019, Mandal2021,Delplace2021,Sayyad2022}. Initial experiments have been carried out in acoustics~\cite{Ding2016,Tang2020,Fang2021,Tang2022}, photonics~\cite{Hodaei2017,Pino2022}, electronic circuits~\cite{Hu2023} and with single photon~\cite{Wang2023}. Notably, the observation of an EX in three-state acoustic systems with parity-time ($\mathcal{PT}$) symmetry was reported~\cite{Tang2020}. However, higher-order exceptional phenomena have yet to be experimentally studied with ultracold atoms. In particular, the atomic realization of an EX is highly desired, as it can pave the way for exploring exotic properties of EX in the many-body setting of quantum gases. 

Here, we experimentally realize an EX based on dissipative Bose-Einstein condensates (BEC) of $^{87}\textrm{Rb}$ atoms. A crucial novelty of our platform is that tunable dissipation exhibits prominent density dependence, arising from the collective response of atoms to resonant light. We implement a dissipative three-state model that hosts an EX within a two-parameter space, despite the absence of symmetry. We measure the system's decay dynamics and analyze the consequence of density-dependent dissipation. Based on an effective description of the transient dynamics, we identify two EAs with different EP geometries, and demonstrate their coalescence produces an EX. These intriguing EAs and EX result from the different roles played by dissipation in the strong coupling limit and quantum Zeno (QZ) regime~\cite{Facchi2008}. Our work differs from prior experiments~\cite{Hodaei2017,Ding2022,Wang2023} where the implementation (i) concerns single-particle dissipation, and (ii) requires at least $3$ degrees of freedom, or uses symmetries to alleviate constraints, often resulting in EAs with identical geometries. Note that versatile EP geometries have recently attracted significant interest~\cite {Ding2022}.

We begin with theoretically describing the physics of a three-level system modeled by the non-Hermitian Hamiltonian (in the basis $|1\rangle, |2\rangle, |3\rangle$)
\begin{eqnarray}
H= \Omega_1\left(\begin{array}{ccc}0&1& 0\\ 1&0& w\\0 &w&-i\bar{\Gamma}\\ \end{array}\right), \label{H}
\end{eqnarray}
where the coherent coupling rate $\Omega_{1}$ between states $|1\rangle$ and $|2\rangle$ is used as the energy unit. Hamiltonian $H$ involves two degrees of freedom: the coherent coupling rate $w\in \mathcal{R}$ between $|2\rangle$ and $|3\rangle$, and the dissipation rate $\bar{\Gamma}\in \mathcal{R}$ of $|3\rangle$. 

\begin{figure}[tp]
\vspace{-0.7cm}
 \begin{center}  
\includegraphics[width=1\columnwidth]{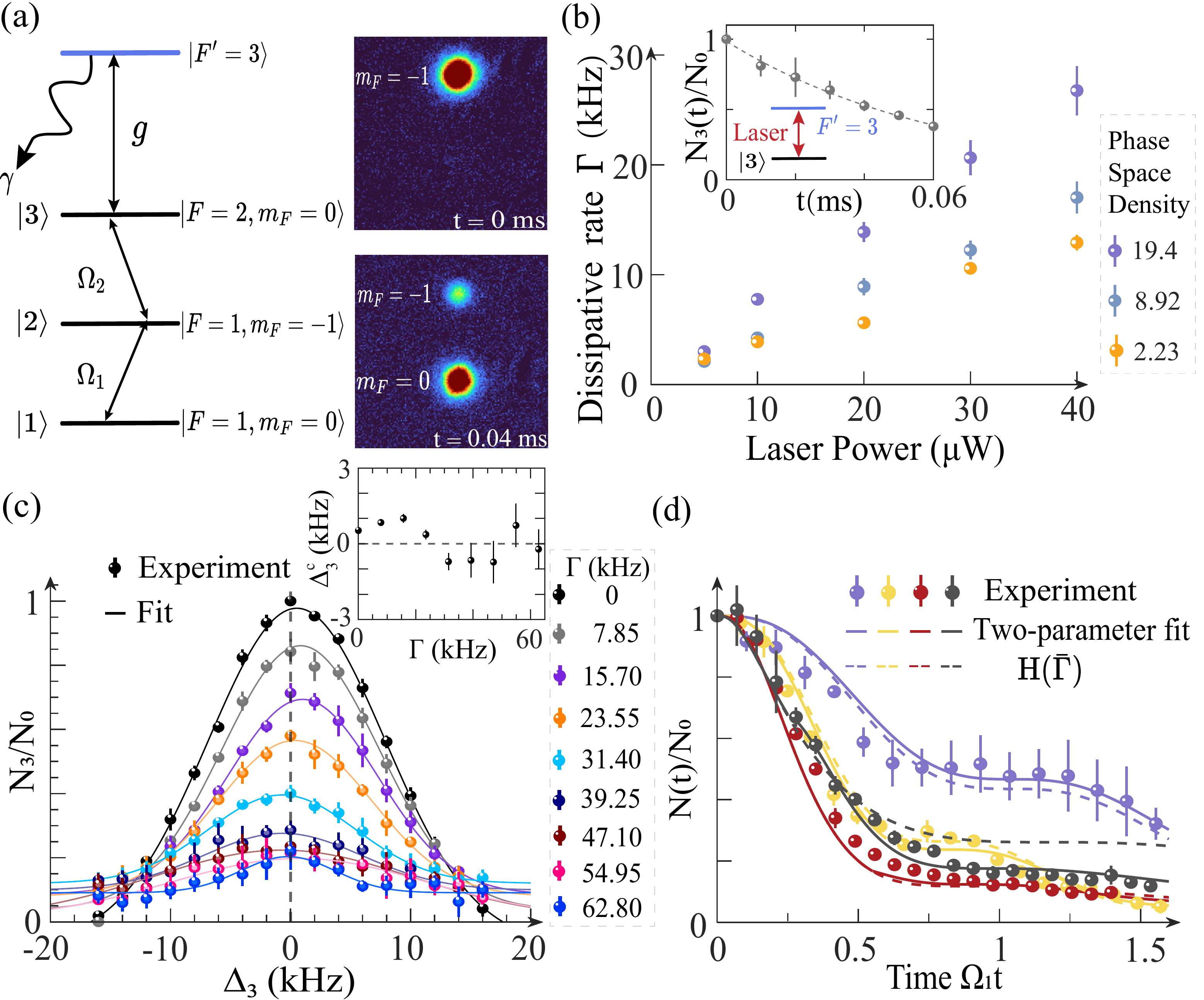}  
\caption{Implementation with BECs of $^{87}\textrm{Rb}$ atoms. (a) Energy level diagram. Three hyperfine states in the ground-state manifold, $|F=1,m_F=0\rangle$, $|F=1,m_F=-1\rangle$ and $|F=2,m_F=0\rangle$, are used to encode $|1\rangle$, $|2\rangle$ and $|3\rangle$. A resonant radio-frequency field couples $|F=1,m_F=0\rangle$ and $|F=1,m_F=-1\rangle$ with the coupling rate $\Omega_1$. A microwave field resonantly couples $|F=1,m_F=-1\rangle$ and $|F=2,m_F=0\rangle$ with the coupling rate $\Omega_2$. A resonant laser drives a transition between $|F=2,m_F=0\rangle$ and an electronically excited state $|F^{\prime}=3\rangle$ with spontaneous emission rate $\gamma=2\pi\times 6.06$ MHz (out of system); the coupling rate $g$ is controlled via laser power. The BEC is initialized in $|F=1,m_F=-1\rangle$. After an evolution time $t$, the Stern-Gerlach absorption image is taken after $10$ ms time of flight. Illustrated images are taken at $t=0$ and $t=0.04$ ms when $w=\Omega_2/\Omega_1=2.8$, with $\Omega_1=10.36$ kHz and $\Omega_2=29.01$ kHz.  (b) Measured dissipation rate $\Gamma$ as a function of laser intensity, for various phase space densities. Inset: Though a fit (dotted curve) of the short-time population dynamics in $|F=2,m_F=0\rangle$ as $N_3(t)/N_0=e^{-2\Gamma t}$, one obtains $\Gamma$ for the phase space density of $19.4$ and the laser power of $10 \mu W$.  (c) Measurement of transfer efficiency from $|F=1, m_F=-1\rangle$ to $|F=2, m_F=0\rangle$ under a wide range of $\Gamma$, when $\Omega_2=31.41$ kHz. By fitting the data (solid curve), the center frequency $\Delta_3^c$ is extracted and shown as a function of $\Gamma$ in the inset. (d) Measured total atom number $N(t)/N_0$ (normalized to the initial number $N_0$) as a function of (dimensionless) time. Purple, yellow, red, and black dots denote the data for $(w=2.8,\bar{\Gamma}=1)$, $(w=3.8,\bar{\Gamma}=2)$, $(w=4.5,\bar{\Gamma}=8)$, $(w=4.5,\bar{\Gamma}=17)$, respectively, with $\bar{\Gamma}=\Gamma/\Omega_1$. Dashed curves denote simulations using $H(\bar{\Gamma})$. Solid curves denote two-parameter fit to the data using $H(\bar{\Gamma}_1)$ and $H(\bar{\Gamma}_2)$; see text and Supplementary Materials. Each data is the average of (b) 5 (c) 5 and (d) 3 measurements. The error bars are $1\sigma$ standard deviations. } 
\label{Fig2}
 \end{center}  
\vspace{-0.7cm}
\end{figure}

Depending on the dissipation $\bar{\Gamma}$ being weak or strong relative to $w$, two EP2s can emerge and eventually lead to EX, which can be qualitatively understood as follows. (i) In the strong-coupling limit $w\gg1$ and $\bar{\Gamma}\lesssim w$, $|1\rangle$ is essentially decoupled. In the subspace spanned by $|2\rangle$ and $|3\rangle$, therefore, the effective Hamiltonian is $H_{\textrm{eff}}/\Omega_1=-i({\bar{\Gamma}}/{2})I+[w\sigma_x+i({\bar{\Gamma}}/{2})\sigma_z]$ with the identity matrix $I$ and Pauli matrices $\sigma_i$ $(i=x,y,z)$, which hosts an EP2 
 \begin{equation}
w_{e_1}-\frac{1}{2}\bar{\Gamma}_{e_1}=0. \label{eq:EP2_1}
\end{equation}
When $\bar{\Gamma}<\bar{\Gamma}_{e_1}$, the $H_{\textrm{eff}}/\Omega_1$ has complex eigenvalues $(\pm\sqrt{w^2-\bar{\Gamma}^2/4})-i\bar{\Gamma}/2$, and when $\bar{\Gamma}>\bar{\Gamma}_{e_1}$, two imaginary eigenvalues $i(-\bar{\Gamma}/2\pm\sqrt{\bar{\Gamma}^2/4-w^2})$ exist. (ii) In the opposite strong-dissipation limit $\bar{\Gamma}\gg w, 1$, the QZ effect~\cite{Facchi2008} occurs and confine the system in the subspace spanned by $|1\rangle$ and $|2\rangle$, decoupled from $|3\rangle$. In this Zeno-subspace, the effective Hamiltonian is $H_{\textrm{eff}}'/\Omega_1=-i({w^2}/{2\bar{\Gamma}})I+[\sigma_x+i({w^2}/{2\bar{\Gamma}})\sigma_z]$. Thus an EP2 occurs at
\begin{equation}
\frac{1}{2}w^2_{e_2}-\bar{\Gamma}_{e_2}=0. \label{eq:EP2_2}
\end{equation}
When $\bar{\Gamma}<\bar{\Gamma}_{e_2}$, the eigenvalues of $H_{\textrm{eff}}'/\Omega_1$ are imaginary and bifurcate. But when $\bar{\Gamma}>\bar{\Gamma}_{e_2}$, the eigen-decay rates $\sim w^2/(2\bar{\Gamma})$ are degenerate, while the real parts bifurcate and asymptotically approach $\pm 1$ for $\bar{\Gamma}\rightarrow \infty$. Different from the linear EP arc in Eq.~(\ref{eq:EP2_1}), Equation~(\ref{eq:EP2_2}) implies a parabola. Thus, increasing $\bar{\Gamma}/w$ in Eq.~(\ref{H}) from $\bar{\Gamma}/w\rightarrow 0$ to $\bar{\Gamma}/w\rightarrow \infty$ may result in two EP2s, whose coalescence creates an EX (EP3). 

The above analysis is numerically verified in Fig.~\ref{Fig1}(a) and Fig.~\ref{Fig1}(b) in the $(\bar{\Gamma}, w)$ plane. The EX occurs at $w= 2\sqrt{2}$ and $\bar{\Gamma}= 3\sqrt{3}$, where eigenvalues are degenerate at $E_{\textrm{EP3}}/\Omega_1= -\sqrt{3}i$, and eigenstates coalesce into $
\left | \psi \right \rangle_\textrm{EP3}=\frac{i}{\sqrt{6} }|1\rangle+\frac{1}{\sqrt{2}}|2\rangle-\frac{i}{\sqrt{3}}|3\rangle$. The different EP geometries of the two arcs in Fig.~\ref{Fig1}(b) agree with Eqs.~(\ref{eq:EP2_1}) and (\ref{eq:EP2_2})~\cite{sup}. 

At the EX, the system exhibits anisotropic responses to perturbations $H_1$~\cite{Demange2012,Xiao2019,Tang2020} that reflects hybrid topological nature of the EX~\cite{Xiao2019,Tang2020}. Consider the perturbed Hamiltonian $H'=H_\textrm{EX}+z H_1$ with $z=\epsilon e^{i\theta}$ ($\epsilon/\Omega_1\rightarrow 0$). Its eigenstates (eigenvalues) exhibit two distinct scaling behaviors [Fig.~\ref{Fig1}(c)]: (i) For perturbations such as $H_1= |3\rangle \langle3 |$, all perturbed  eigenstates (eigenvalues) scale as $\propto z^{1/3}$; (ii) For perturbations like $H_1=i|1\rangle\langle 3|-i|3\rangle\langle 1|+\frac{\sqrt{3}}{3}|2\rangle\langle 1|+\frac{\sqrt{6}}{6}|2\rangle\langle 3|$, two eigenstates (eigenvalues) scale as $\propto z^{1/2}$, while the remaining one $\propto z$\cite{sup}. To reveal topological property of the states near the EX, we calculate~\cite{sup} the Berry phase $\phi_B$ accumulated in the parallel transport~\cite{Xiao2019,Tang2020,Soluyanov2012} of a perturbed eigenstate in a closed loop depicted in Fig.~\ref{Fig1}(c). For $z^{1/3}$-perturbation, we find that all eigenstates return to their initial point after three cycles and accumulate a quantized Berry phase of $\phi_B=-2\pi$, same as $\mathcal{PT}$-symmetric scenarios. In case (ii), however, we find that two eigenstates, which scale as $z^{1/2}$, return after two cycles, with $\phi_B=-2.17\pi$, while the remaining eigenstate returns after just one cycle, with $\phi_B=-0.83\pi$; these Berry phases are not quantized due to the absence of symmetry~\cite{Mailybaev2005}, unlike $\mathcal{PT}$-symmetric scenarios~\cite{Xiao2019,Tang2020}.

Experimentally, we implement Eq.~(\ref{H}) based on the $^{87}\textrm{Rb}$ BEC as shown in Fig.~\ref{Fig2}. We exploit three ground-state hyperfine states ($|F=1,m_F=0\rangle$, $|F=1,m_F=-1\rangle$ and $|F=2,m_F=0\rangle$) of $^{87}\textrm{Rb}$ atoms, respectively, to encode $|1\rangle$, $|2\rangle$ and $|3\rangle$. Level couplings are shown in Fig.~\ref{Fig2}(a). To realize tunable dissipations in $|3\rangle$, we use a laser light at $\lambda=780$ nm to drive a resonant transition from $|F=2,m_F=0\rangle$ to an electronically excited state $|F^{\prime}=3\rangle$ with atomic linewidth $\gamma=2\pi\times 6.06$ MHz, such that when atoms populate $|F=2,m_F=0\rangle$, they are lost from the system at a rate $\Gamma$. Tunable $\Gamma$ is achieved by controlling the laser power. In all measurements~\cite{sup}, we start with preparing a BEC with a number $N_0\sim 2.5 \times10^4$ $^{87}\textrm{Rb}$ atoms in $|F=1,m_F=-1\rangle$ at the temperature $\sim 50$ nK. 

\begin{figure*}[t]
\vspace{-0.5cm}    
 \begin{center}  
\includegraphics[width=0.9\textwidth]{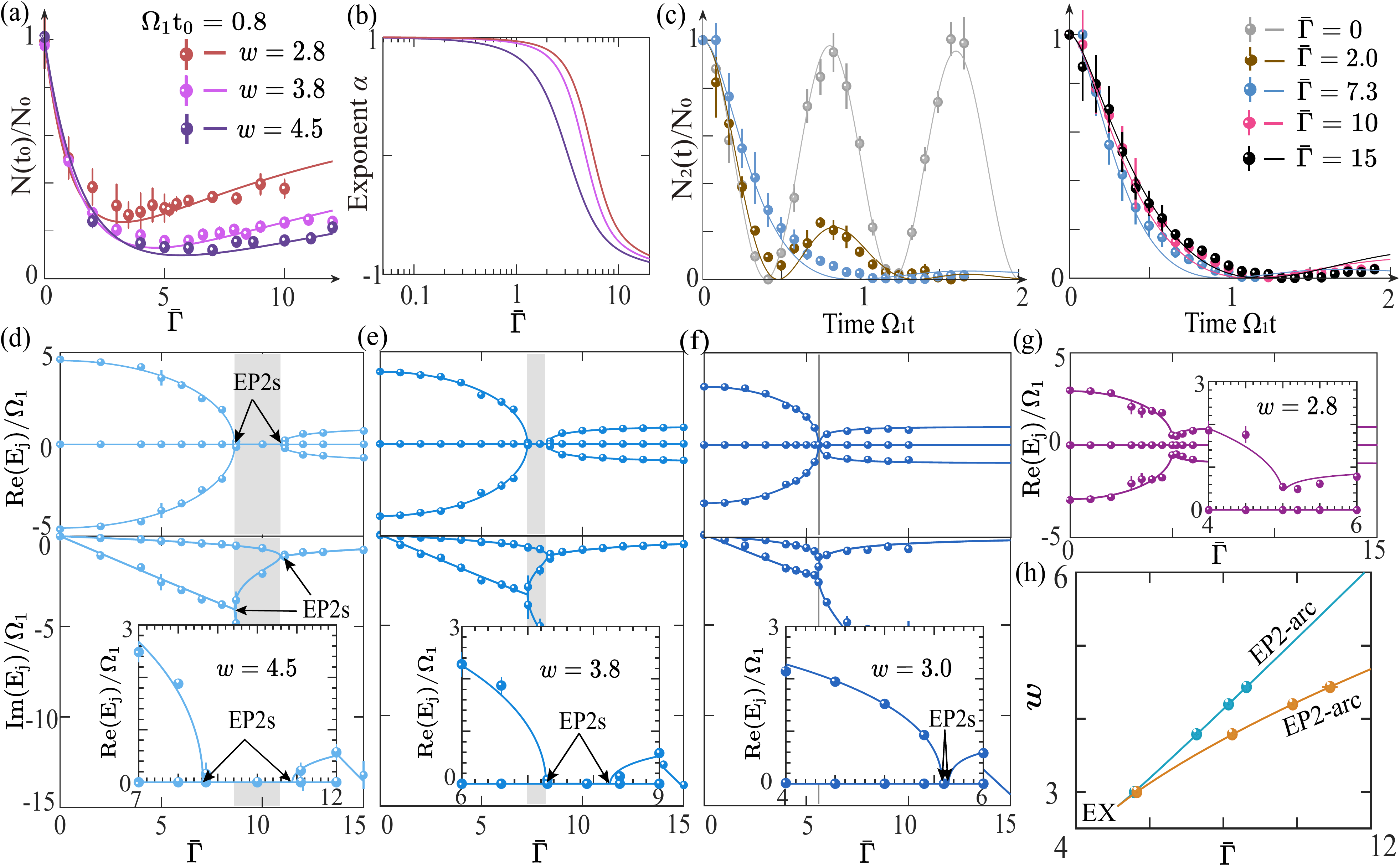}  
 \caption{Measurement of EAs and EX. (a) Measured total atom number $N(t_0)/N_0$ at time $\Omega_1t_0=0.8$ as a function of the dissipation rate $\bar{\Gamma}$. The experimental data are shown for the coupling rate $w=2.8,3.8,4.5$~\cite{sup}, respectively. Data within the time window $\Omega_1t_0=0.8\pm 0.05$ are shown. The solid curves are the numerical fittings of the experimental data as a function of $\bar{\Gamma}$. (b) Exponent $\alpha$ as a function of $\bar{\Gamma}$. Assuming the fitted $N(t_0)/N_0$ in (a) is of the form $N(t_0)/N_0 \simeq \exp(-\gamma t_0)$ with the scaling $\gamma\propto \bar{\Gamma}^\alpha$, we extract the exponent $\alpha$ as a function of $\bar{\Gamma}$, for $w=2.8,3.8,4.5$ (red, pink and purple), respectively. (c) Measured time evolution of the atom number $N_2(t)/N_0$ in $|F=1,m_F=-1\rangle$, for $w=3.8$. Left panel: $\bar{\Gamma}/w\lesssim 1$. Right panel: $\bar{\Gamma}/w\gtrsim 1$. Solid curves are two-parameter fit to the data~\cite{sup}. (d)-(g) Measured eigenvalues as a function of $\bar{\Gamma}$, when (d) $w=4.5$, (e) $w=3.8$, (f) $w=3$ and (g) $w=2.8$. In (d)-(f), both real (top) and imaginary (bottom) parts of the eigenvalues are shown. The gray shaded area indicates regions between the two EP2s. In (g), where $w=2.8<w_\textrm{EX}$, only the real part of eigenvalues is shown. The experimental data of eigenvalues are obtained through the fitted $H(\bar{\Gamma}_1)$ (see text). By further fitting (solid curves) the measured eigenvalues as a function of $\bar{\Gamma}$, we identify the EPs. Insets of (d)-(f) are zoom-in plots of the measured eigenvalues near EPs. (h) Experimental EAs and EX in a two-dimensional parameter space $(\bar{\Gamma},w)$ shown on top of the theoretical results (solid curves). In (a) and (c), each data is the average over 3 measurements, and error bars denote $1\sigma$ standard deviations. In (d)-(h), error bars denote the fitting errors.}  \label{Fig3}
 \end{center} 
\vspace{-0.7cm}
\end{figure*}

To determine the dissipation rate $\Gamma$ [inset of Fig.~\ref{Fig2}(b)], we use microwave (MW) $\pi$ pulse to transfer the BEC from $\left | F=1,m_F=-1  \right \rangle $ to $\left | F=2,m_F=0  \right \rangle $; subsequently, the MW field is switched off while the resonant laser is switched on. By monitoring the short-time dynamics of the population $N_3(t)$ in $\left | F=2,m_F=0  \right \rangle $, we obtain $\Gamma$ through an appropriate fit of data via the formula $N_3(t)=N_0e^{-2\Gamma t}$. We have checked that, in the absence of the dissipation beam, the natural atom loss in the BEC is negligible over times $\sim 10$ ms~\cite{sup}. 

Given that the BEC has a size comparable to the light wavelength $\lambda$ and exhibits a dense atomic density $n\sim 3.4\times 10^{13}\textrm{cm}^{-3}$, we estimate $n\lambda^3\sim 16\gg 1$ in a regime characterized by the emergence of \textit{collective} optical response of the atomic cloud to a resonant light~\cite{You1994,Pellegrino2014,Jenkins2016,Schilder2020}. Consequently, we observe a collective, density-dependent dissipation rate $\Gamma(n)$ arises, distinct from noninteracting scenarios. Figure~\ref{Fig2}(b) illustrates the measured $\Gamma$ as a function of the excitation laser power, for various phase space density of atoms, below and above the critical phase space density $\sim 2.6$ for condensation. We observe enhanced dissipation for increased density, particularly at higher laser intensities, which indicates modifications to the single-particle dissipation rate $g^2/\gamma$. A comprehensive study of this light-induced many-body dissipation of BECs will be presented in a separate work. To check if the dissipation beam causes any resonance shift, we measure the transfer efficiency from $|F=1, m_F=-1\rangle$ to $|F=2, m_F=0\rangle$ sweeping the detuning $\Delta_3$ of $|F=2, m_F=0\rangle$, for a wide range of $\Gamma$ [Fig.~\ref{Fig2}(c)]. After extracting the center frequency (inset), we find no visible systematic shift. Below, we denote the measured dissipation rate as $\bar{\Gamma}=\Gamma/\Omega_1$.

The density-dependent dissipation rate in the implemented three-state system gives rise to intriguing decay dynamics over a relatively long time scale of $\sim 0.2$ ms. Experimentally, after preparing a BEC in $|F=1,m_F=-1\rangle$, we sweep the bias magnetic field to 16 G within $100$ ms and wait $100$ ms for the bias field to stabilize. Once stabilized, we abruptly turn on the MW field, RF field, and the resonant laser~\cite{sup} to drive the time evolution of the BEC. After an evolution time $t$, we utilize the Stern-Gerlach absorption imaging technique to measure the atomic populations in the two Zeeman levels [Fig.~\ref{Fig2}(a)], respectively. Figure~\ref{Fig2}(d) shows the measured (normalized) total atom number $N(t)/N_0$, for various $w$ and $\bar{\Gamma}$~\cite{sup}. The dashed curves denote simulations using Hamiltonian~(\ref{H}) with $\bar{\Gamma}$. The simulation closely matches the experiment at transient times. Nevertheless, discrepancies may arise after some time, particularly as $\bar{\Gamma}$ increases. This deviation can be understood since the atomic cloud becomes dilute over time, so that the density-dependent dissipation rates at later times may differ considerably from that at initial stages. Further comparisons between the experiment and the simulation with $H(\bar{\Gamma})$ (dashed curves) seem to suggest decreased dissipation rates $<\bar{\Gamma}$ at long times, as illustrated, for instance, by the black curves representing the Zeno limit.  

Motivated by the observations that $H({\bar{\Gamma}})$ remains a good description of the system dynamics at short time scales, while long-time dynamics may exhibit decreased dissipation rates, we adopt a minimal strategy to effectively model the full dynamics, which is referred to as the two-parameter fit below. The approach consists of using two Hamiltonians of the form (\ref{H}), $H(\bar{\Gamma}_{1})$ and $H(\bar{\Gamma}_{2})$, to model the dynamics at times $0\le t<t_m$ and $t\ge t_m$, respectively, with three parameters, $\bar{\Gamma}_{1}$, $\bar{\Gamma}_{2}$ and $t_m$, to be determined from the best fit to the data. In the fitting process, we first choose the initial values of $\bar{\Gamma}_{1,2}$ and $t_m$ according to the following considerations: For $\bar{\Gamma}_{1}$, we take its initial value to be $\sim\bar{\Gamma}$, because $H({\bar{\Gamma}})$ offers a good short-time description as mentioned earlier; For $t_m$, we take its initial value to be the instant when the experimental data is observed to deviate notably from the model with $H(\bar{\Gamma})$; For $\bar{\Gamma}_{2}$, we choose its initial value to be $\sim \bar{\Gamma}/2$. Then, starting from these initial values, we iteratively improve the fitting parameters until the difference of the calculated $N(t)/N_0$ and the measured dynamics are minimized in the least-square sense. As shown in Fig.~2(d) and detailed in~\cite{sup}, such a two-parameter fit leads to a much better agreement with the experiment compared to utilizing a single dissipation rate. 

Despite complications caused by the density dependence of the dissipation rate, we can probe some generic features associated with the different interplays of dissipation and coherent processes in the weak and strong dissipation limits as expected from the model (\ref{H}). In Fig.~\ref{Fig3}(a), the measured $N(t_0)/N_0$ at some fixed time $\Omega_1t_0=0.8\pm 0.05$ is shown as a function of $\bar{\Gamma}$~\cite{sup}. For $\bar{\Gamma}/w<1$, $N(t_0)/N_0$ decreases with $\bar{\Gamma}$, indicating enhanced atomic loss; the curves with different $w$ significantly overlap, which signals that the underlying eigen-decay rates are insensitive to $w$. These features agree with previous predictions in the strong coupling limit (i). When $\bar{\Gamma}/w\gg1$, in contrast, we observe a revival of population when $\bar{\Gamma}$ increases, in such a way that depends on $w$, consistent with previous analysis in Zeno regime (ii). To further reveal how the atom loss behaves under different $\bar{\Gamma}$, we fit the measured $N(t_0)/N_0$ as a function of $\bar{\Gamma}$ (solid curve) and assume $N(t_0)/N_0 \simeq \exp(-\gamma t_0)$ with $\gamma\propto \bar{\Gamma}^\alpha$. We extract the exponent $\alpha$ as a function of $\bar{\Gamma}$ in Fig.~\ref{Fig3}(b). The $\alpha$ under various $w$ exhibits similar asymptotic behaviors, turning from $\alpha\simeq1$ when $\bar{\Gamma}/w\rightarrow 0$ toward $\alpha\rightarrow -1$ when $\bar{\Gamma}/w\gg1$, as expected in this two limits. We also measure the atom population $N_2(t)/N_0$ in $|F=1,m_F=-1\rangle$ [Fig.~\ref{Fig3}(c)]. In the left panel with $\bar{\Gamma}\lesssim w$, clear oscillation is observed, but the oscillation frequency is reduced with $\bar{\Gamma}$. This indicates the underlying energy spectrum is non-degenerate and decreases with $\bar{\Gamma}$. The oscillation seems to revive at strong dissipations $\bar{\Gamma}\gtrsim w$ [right panel of Fig.~\ref{Fig3}(c)]. But the full oscillation cannot be observed within the measurement time scale, which needs more sophisticated techniques~\cite{Trypogeorgos2018}. 

Finally, we probe the existence of EX, which is not only an EP3 but also the cusp singularity of EAs, using the Hamiltonian $H(\bar{\Gamma}_1)$ that appropriately yields the short-time system dynamics. The  measured eigenvalues are shown in Figs.~\ref{Fig3}(d)-(g) as a function of the nominal $\bar{\Gamma}$, for $w=4.5,3.8,3, 2.8$, respectively. By fitting the measured eigenvalues as a function of $\bar{\Gamma}$, we identify potential EPs as the degenerate points. In Fig.~\ref{Fig3}(d)-(f), we observe two EP2s that gradually merge when reducing $w$. In Fig.~\ref{Fig3}(f), the two sufficiently close EP2s strongly evidence the existence of an EP3. When further reducing $w$ below EP3 as in Fig.~\ref{Fig3}(g), all three eigenvalues become non-degenerate again. This signals the trajectories of EP2s terminate at EP3, i.e., to form the cusp singularity. Finally, we collect the experimental data of EPs to construct their trajectories in the $(\bar{\Gamma},w)$ plane [Fig.~\ref{Fig3}(h)]. The good agreement between the experiment and theory (solid curves) evidences the two EAs with different EP geometries, whose coalescence leads to an EX.

Summarizing, using dissipative BECs, we realize an EX formed by the coalescence of EAs with different EP geometries, despite the absence of symmetry. In our implementation, tunable dissipation has a many-body nature, and features density dependence. This feature makes dissipative BECs under resonant light a unique platform in the study of non-Hermitian physics.

{\it Acknowledgements} -- We acknowledge discussions with Mingyong Jing. This research is funded by the National Key Research and Development Program of China (Grants No. 2022YFA1404201, No. 2022YFA1203903, No. 2022YFA1404003), the National Natural Science Foundation of China (Grants No. 12274272, No. 12374246, No. 12034012, No. 12074231), and NSFC-ISF (No. 12161141018). Y.H. acknowledges support by Beijing National Laboratory for Condensed Matter Physics (No. 2023BNLCMPKF001).


\begin{thebibliography}{99}

\bibitem{Heiss2012} W. Heiss, The Physics of Exceptional Points, J. Phys. A: Math. Theor. \textbf{45}, 444016 (2012).

\bibitem{Ashida2020} Y. Ashida, Z. Gong, and M. Ueda, Non-Hermitian Physics, Adv. Phys. \textbf{69}, 249 (2020).

\bibitem{Bergholtz2021} E. Bergholtz, J. Budich, and F. Kunst, Exceptional Topology of Non-Hermitian Systems, Rev. Mod. Phys. \textbf{93}, 015005 (2021).

\bibitem{Ding2022} K. Ding, C. Fang, and G. Ma, Non-Hermitian Topology and Exceptional-Point Geometries, Nat. Rev. Phys. \textbf{4}, 745 (2022).

		

\bibitem{Peng2016} P. Peng, W. Cao, C. Shen, W. Qu, J. Wen, L. Jiang, and Y. Xiao, Anti-Parity–Time Symmetry with Flying Atoms, Nat. Phys. \textbf{12}, 1139 (2016).

\bibitem{Li2019} J. Li, A. Harter, J. Liu, L. Melo, Y. Joglekar, and L. Luo, Observation of Parity-Time Symmetry Breaking Transitions in a Dissipative Floquet System of Ultracold Atoms, Nat. Commun. \textbf{10}, 855 (2019).

\bibitem{Ren2022} Z. Ren, D. Liu, E. Zhao, C. He, K. Pak, J. Li, and G. Jo, Chiral Control of Quantum States in Non-Hermitian Spin–Orbit-Coupled Fermions, Nat. Phys. \textbf{18}, 385 (2022).

\bibitem{Liang2023} C. Liang, Y. Tang, A. Xu, and Y. Liu, Observation of Exceptional Points in Thermal Atomic Ensembles, Phys. Rev. Lett. \textbf{130}, 263601 (2023).

\bibitem{DingL2021}L. Ding, K. Shi, Q. Zhang, D. Shen, X. Zhang, and W. Zhang, Experimental Determination of PT-Symmetric Exceptional Points in a Single Trapped Ion, Phys. Rev. Lett. \textbf{126}, 083604 (2021).
   
 
\bibitem{Wu2019} Y. Wu, W. Liu, J. Geng, X. Song, X. Ye, C. Duan, X. Rong, and J. Du, Observation of Parity-Time Symmetry Breaking in a Single-Spin System, Science \textbf{364}, 878 (2019).

\bibitem{Choi2010} Y. Choi, S. Kang, S. Lim, W. Kim, J. Kim, J. Lee, and K. An, Quasieigenstate Coalescence in an Atom-Cavity Quantum Composite, Phys. Rev. Lett. \textbf{104}, 153601 (2010).


\bibitem{Ding2016} K. Ding, G. Ma, M. Xiao, Z. Zhang, and C. Chan, Emergence, Coalescence, and Topological Properties of Multiple Exceptional Points and Their Experimental Realization, Phys. Rev. X \textbf{6}, 021007 (2016).

\bibitem{Graefe2008} E. Graefe, U. Günther, H. Korsch, and A. Niederle, A Non-Hermitian $\mathcal{P}\mathcal{T}$ Symmetric Bose–Hubbard Model: Eigenvalue Rings from Unfolding Higher-Order Exceptional Points, J. Phys. A: Math. Theor. \textbf{41}, 255206 (2008).

\bibitem{Demange2012} G. Demange and E. Graefe, Signatures of Three Coalescing Eigenfunctions, J. Phys. A: Math. Theor. \textbf{45}, 025303 (2012).

\bibitem{Xiao2019} Y. Xiao, Z. Zhang, Z. Hang, and C. Chan, Anisotropic Exceptional Points of Arbitrary Order, Phys. Rev. B \textbf{99}, 241403(R) (2019).

\bibitem{Tang2020} W. Tang, X. Jiang, K. Ding, Y. Xiao, Z. Zhang, C. Chan, and G. Ma, Exceptional Nexus with a Hybrid Topological Invariant, Science \textbf{370}, 1077 (2020).


\bibitem{Mandal2021} I. Mandal and E. J. Bergholtz, Symmetry and Higher-Order Exceptional Points, Phys. Rev. Lett. \textbf{127}, 186601 (2021).

\bibitem{Delplace2021} P. Delplace, T. Yoshida, and Y. Hatsugai, Symmetry-Protected Multifold Exceptional Points and Their Topological Characterization, Phys. Rev. Lett.  \textbf{127}, 186602 (2021).


\bibitem{Sayyad2022} S. Sayyad and F. Kunst, Realizing Exceptional Points of Any Order in the Presence of Symmetry, Phys. Rev. Res.  \textbf{4}, 023130 (2022).


\bibitem{Pan2019} L. Pan, S. Chen, and X. Cui, High-Order Exceptional Points in Ultracold Bose Gases, Phys. Rev. A \textbf{99}, 011601(R) (2019).


\bibitem{Fang2021} X. Fang, N. Gerard, Z. Zhou, H. Ding, N. Wang, B. Jia, Y. Deng, X. Wang, Y. Jing, and Y. Li, Observation of Higher-Order Exceptional Points in a Non-Local Acoustic Metagrating, Commun. Phys. \textbf{4}, 271 (2021).


\bibitem{Tang2022} W. Tang, K. Ding, and G. Ma, Experimental Realization of Non-Abelian Permutations in a Three-State Non-Hermitian System, Natl. Sci. Rev. \textbf{9}, nwac010 (2022).


\bibitem{Hodaei2017} H. Hodaei, A. Hassan, S. Wittek, H. Garcia-Gracia, R. Ganainy, D. Christodoulides, and M. Khajavikhan, Enhanced Sensitivity at Higher-Order Exceptional Points, Nature \textbf{548}, 187 (2017).


\bibitem{Pino2022} J. Pino, J. Slim, and E. Verhagen, Non-Hermitian Chiral Phononics through Optomechanically Induced Squeezing, Nature \textbf{606}, 82 (2022).

\bibitem{Hu2023} J. Hu, R. Zhang, Y. Wang, X. Ouyang, Y. Zhu, H. Jia, and C. Chan, Non-Hermitian Swallowtail Catastrophe Revealing Transitions among Diverse Topological Singularities, Nat. Phys. \textbf{19}, 1098  (2023).

\bibitem{Wang2023} K. Wang, L. Xiao, H. Lin, W. Yi, E. Bergholtz, and P. Xue, Experimental Simulation of Symmetry-Protected Higher-Order Exceptional Points with Single Photons, Sci. Adv. \textbf{9}, eadi0732 (2023).

\bibitem{Facchi2008} {P. Facchi and S. Pascazio, Quantum Zeno Dynamics: Mathematical and Physical Aspects, J. Phys. A: Math. Theor. \textbf{41}, {493001} (2008).}


\bibitem{sup} See details in Supplementary Materials, which include Refs.~\cite{Demange2012,Xiao2019,Soluyanov2012,Wang2021,Hu2021,Patil2022,Hu2022,Guo2023}. There, we provide details on the exact solutions of eigenvalues, experimental details on the calibration and fitting, supplementary data on the dynamics, and detailed theory on the anisotropic perturbation behaviors around the EX (including anisotropic eigenvalues and hybrid topological property).


\bibitem{Soluyanov2012} A. Soluyanov and D. Vanderbilt, Smooth Gauge for Topological Insulators, Phys. Rev. B \textbf{85}, 115415 (2012).


\bibitem{Mailybaev2005} A. Mailybaev, O. Kirillov, and A. Seyranian, Geometric Phase around Exceptional Points, Phys. Rev. A \textbf{72}, 014104 (2005).

\bibitem{You1994} L. You, M. Lewenstein, and J. Cooper, Line Shapes for Light Scattered from Bose-Einstein Condensates, Phys. Rev. A \textbf{50}, R3565 (1994).

\bibitem{Pellegrino2014} J. Pellegrino, R. Bourgain, S. Jennewein, Y. Sortais, A. Browaeys, S. Jenkins, and J. Ruostekoski, Observation of Suppression of Light Scattering Induced by Dipole-Dipole Interactions in a Cold-Atom Ensemble, Phys. Rev. Lett. \textbf{113}, 133602 (2014).

\bibitem{Jenkins2016} S. Jenkins, J. Ruostekoski, J. Javanainen, S. Jennewein, R. Bourgain, J. Pellegrino, Y. Sortais, and A. Browaeys, Collective Resonance Fluorescence in Small and Dense Atom Clouds: Comparison between Theory and Experiment, Phys. Rev. A \textbf{94}, 023842 (2016).

\bibitem{Schilder2020} N. Schilder, C. Sauvan, Y. Sortais, A. Browaeys, and J. Greffet, Near-Resonant Light Scattering by a Subwavelength Ensemble of Identical Atoms, Phys. Rev. Lett. \textbf{124}, 073403 (2020).

\bibitem{Trypogeorgos2018}  D. Trypogeorgos, A. Valdés-Curiel, I. Spielman, and C. Emary, Perpetual Emulation Threshold of $\mathcal{PT}$ -Symmetric Hamiltonians, J. Phys. A: Math. Theor. \textbf{51}, 325302 (2018).

\bibitem{Wang2021} K. Wang, A. Dutt, C. Wojcik, and S. Fan, Topological Complex-Energy Braiding of Non-Hermitian Bands, Nature \textbf{598}, 59 (2021).

\bibitem{Hu2021} H. Hu and E. Zhao, Knots and Non-Hermitian Bloch Bands, Phys. Rev. Lett. \textbf{126}, 010401 (2021).

\bibitem{Patil2022} Y. Patil, J. Höller, P. Henry, C. Guria, Y. Zhang, L. Jiang, N. Kralj, N. Read, and J. Harris, Measuring the Knot of Non-Hermitian Degeneracies and Non-Commuting Braids, Nature \textbf{607}, 271 (2022).

\bibitem{Hu2022} H. Hu, S. Sun, and S. Chen, Knot Topology of Exceptional Point and Non-Hermitian No-Go Theorem, Phys. Rev. Res.  \textbf{4}, L022064 (2022).

\bibitem{Guo2023} C. Guo, S. Chen, K. Ding, and H. Hu, Exceptional Non-Abelian Topology in Multiband Non-Hermitian Systems, Phys. Rev. Lett. \textbf{130}, 157201 (2023).

		
\end{thebibliography}
\end{document}